\documentclass[%
 reprint,
 amsmath,amssymb,
 aps,
]{revtex4-2}

\usepackage{hyperref}
\usepackage{graphicx}
\usepackage{dcolumn}
\usepackage{mathrsfs}
\usepackage{bbold}
\usepackage{bm}
\usepackage[dvipsnames]{xcolor}
\usepackage[normalem]{ulem}

\begin{document}

\newcommand{\ANC}[1]{{\color{blue}{[ANC: #1]}}}

\preprint{APS/123-QED}

\title{Gigahertz-frequency Lamb wave resonator cavities on suspended lithium niobate \newline for quantum acoustics}%


\author{Michele Diego$^{1}$}
\email{diego@iis.u-tokyo.ac.jp}

\author{Hong Qiao$^{2}$}
\thanks{These authors contributed equally to this work: Michele Diego, Hong Qiao}

\author{Byunggi Kim$^{1,3}$}

\author{Minseok Ryu$^{2}$}

\author{Shiheng Li$^4$}

\author{Gustav Andersson$^{2}$}

\author{Masahiro Nomura$^{1}$}
\email{nomura@iis.u-tokyo.ac.jp}

\author{Andrew N. Cleland$^2$}
\email{anc@uchicago.edu}

\affiliation{$^{1}$Institute of Industrial Science, The University of Tokyo, Tokyo 153-8505, Japan\\
$^{2}$Pritzker School of Molecular Engineering, University of Chicago, Chicago, Illinois 60637, USA\\
$^{3}$Department of Mechanical Engineering, School of Engineering, Institute of Science Tokyo, Tokyo 152-8550, Japan\\
$^{4}$Department of Physics, University of Chicago, Chicago IL 60637, USA
}

%
%



\begin{abstract}
Phononic nanodevices offer a promising route toward quantum technologies, as phonons combine strong confinement within matter with broad coupling capabilities to various quantum systems. In particular, the piezoelectric response of materials such as lithium niobate enables coupling between superconducting qubits and gigahertz-frequency phonons. However, bulk lithium niobate phononic devices typically rely on surface acoustic waves and are therefore inherently subject to leakage from the surface into the bulk substrate. Here, we explore the acoustic behavior of resonator cavities supporting GHz-frequency Lamb waves in a 200 nm-thick suspended lithium niobate layer. We characterize the acoustic response at both room and millikelvin temperatures. We find that our resonator cavities with strong confinement reach intrinsic quality factors of approximately 6000 at the single phonon level. We use the measured parameters of the resonators to model their coupling to a superconducting transmon qubit, allowing us to evaluate their potential as quantum acoustic devices.

\end{abstract}

\maketitle


\section{\label{sec:level1}Introduction}

Quantum devices that can control and store quantum states \cite{MacCabe2020, tuokkola2025methods} represent a foundational step toward making quantum computing a real-world application. In this context, phonons have shown great potential as carriers of quantum information \cite{OConnell2010, Chu2017, Satzinger2018, ArrangoizArriola2019}, where one of their main advantages is coupling disparate quantum systems, such as photons \cite{riedinger2016non,mirhosseini2020superconducting}, color centers \cite{teissier2014strain, shandilya2021optomechanical,Whiteley2019}, and superconducting qubits \cite{Gustafsson2014, dumur2021quantum}.
A common method to couple superconducting qubits to acoustic phonons is by combining piezoelectric materials and e.g. interdigitated transducers (IDTs), comprising a set of interleaved metal electrodes patterned on a piezoelectric substrate that convert electrical signals into acoustic waves and vice versa. The emission and detection of the resulting surface acoustic waves on piezoelectric substrates \cite{magnusson2015surface} have been exploited to achieve quantum entanglement \cite{bienfait2020, Wollack2022, Chou2025}, quantum memories \cite{Bozkurt2025}, and to enable the transmission \cite{Bienfait2019} and manipulation \cite{Manenti2017, Qiao2023, Qiao2025a} of quantum information. 

However, devices that employ surface acoustic waves on bulk piezoelectric substrates are intrinsically subject to scattering and energy loss into the bulk \cite{schuetz2015universal}. In seeking to overcome this limitation, one approach is to use a thin suspended layer of piezoelectric material, where the acoustic excitations are known as Lamb waves. To this end, structures featuring IDTs placed on suspended piezoelectric membranes have been implemented using aluminum nitride to make acoustic resonators \cite{chou2020measurements}, as well as lithium niobate for acoustic transmission lines \cite{sarabalis2020s}, Purcell filters \cite{cleland2019mechanical}, tunable transducers \cite{Hugot2026}, along with modeling calculations \cite{arrangoiz2016engineering} and devices for Lamb-mode resonators \cite{ liang2025deployable}.

In this work, we experimentally study suspended lithium niobate resonator cavities incorporating IDTs that are laterally terminated by acoustic Bragg mirrors. We show that the observed resonance corresponds to an antisymmetric $A_0$ Lamb mode at a few GHz. We report measurements of these resonators at both room and millikelvin temperatures, for several choices of IDT–mirror spacings, and we discuss the behavior of the quality factor in the two regimes.
For the cavity with the smallest effective cavity length, we find a high-power quality factor of $Q\approx 6600$, falling somewhat at lower acoustic powers to just above $Q \approx 5300$. We extract from these measurements a lumped-element model for each resonator, and based on these models,  we provide guidance for coupling these resonators to superconducting transmon qubits for quantum-level measurements.

\section{Results and discussion}

\subsection{Room temperature regime}

\begin{figure*}[thb]
\centering
\includegraphics[width=0.925\textwidth]{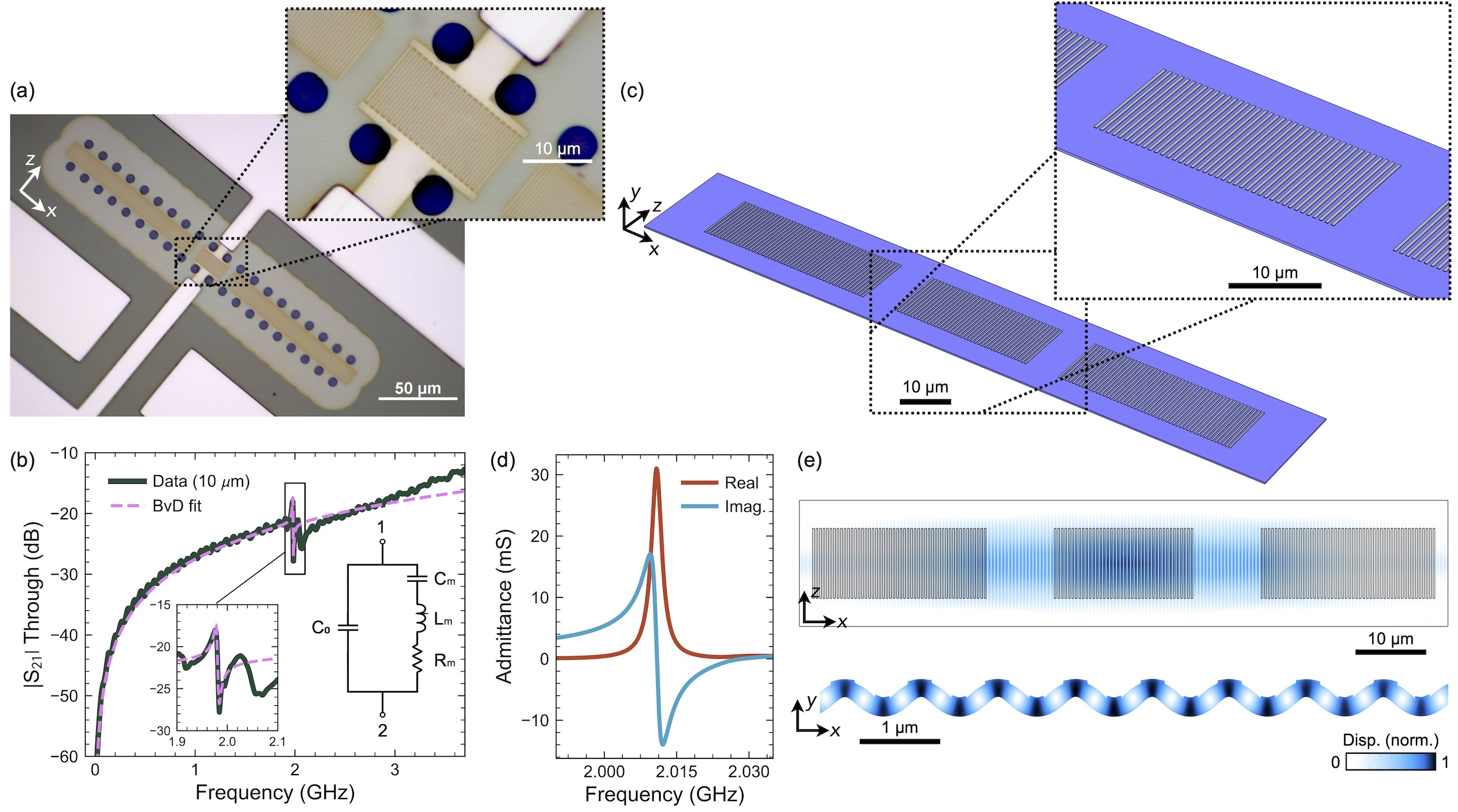}
\caption{\label{fig:1}
Lamb wave resonator cavities on suspended lithium niobate.
(a) Optical microscope image and magnified detail of a representative suspended resonator cavity, consisting of a central IDT terminated along each emission direction by a set of acoustic mirrors, the whole patterned on a suspended 200-nm-thick Y-cut lithium niobate plate. For the cavity here, the IDT–mirror spacing is 5~$\mu$m.
(b) Measured room-temperature $|S_{21}|$ transmission through the IDT of a resonator cavity with an IDT-mirror spacing of 10~$\mu$m. The fit is based on the Butterworth-van Dyke (BvD) model \cite{chou2020measurements}, whose equivalent electrical circuit is shown inset; ports 1 and 2 correspond to the VNA ports used for this measurement.
(c) Model for the resonator cavity used for finite element simulations.
(d) Real and imaginary parts of the simulated admittance, calculated via a finite element simulation for a resonator cavity with an IDT–mirror spacing of 10~$\mu$m. The resonance frequency agrees well with experimental measurements. (e) Displacement field magnitude for the simulated resonant mode, showing the confined antisymmetric $A_0$ Lamb mode from a top-view perspective of the entire cavity (top) and from a cross-sectional view of the cavity center (bottom). 
}
\end{figure*}

The acoustic resonator cavities comprise aluminum IDTs enclosed along the IDT emission directions by acoustic mirrors. Each IDT is composed of 20 equally spaced electrode pairs with a pitch of 500 nm and a metallization ratio of 0.5, resulting in a unit period $p= 1$~$\mu$m.
The electrodes are connected alternately to bond pads forming ports 1 and 2, and driven by a periodic voltage across these ports that excites acoustic waves in the underlying 200 nm thick lithium niobate plate. The acoustic mirrors comprise 200 electrically floating aluminum fingers 250 nm thick. We vary the spacing between the IDT and the mirrors for different devices. Figure~\ref{fig:1}a shows an example of a cavity, where adjacent to the IDT and mirrors, circular windows were etched through the lithium niobate plate to expose the underlying sacrificial silicon dioxide layer, which was later removed to mechanically suspend the resonator cavities (see Appendix \ref{app:A}).  The perimeter of the suspended region can be seen in Fig.~\ref{fig:1}a as a sudden change in the color of the device, where the lighter region corresponds to the suspended area. The devices were suspended by etching in hydrofluoric acid (HF) vapor, which can degrade the IDT and mirror aluminum fingers when etching for more than a few hours. To limit the HF etching time and thus prevent damage to the electrodes, we set the IDT aperture of the cavities to 10~$\mu$m.

Figure~\ref{fig:1}b shows the measured transmission $S_{21}(f)$ through the IDT for a resonator with an IDT–mirror spacing of 10~$\mu$m, measured at room temperature using a two-port vector network analyzer (VNA) and measuring the transmission between ports 1 and 2 as shown inset in this panel \cite{Satzinger2018quantumthesis}. The resonator cavity exhibits the expected resonance at approximately 2 GHz. We fit the spectrum using the Butterworth–van Dyke (BvD) model (inset), which uses a lumped-element equivalent circuit composed of the IDT capacitance $C_0$ in parallel with a branch that includes the equivalent mechanical inductance $L_m$, capacitance $C_m$, and resistance $R_m$ representing mechanical loss. We estimate $C_0 \approx 0.07$ pF, in agreement with modeling calculations.

\begin{figure*}[thb]
\centering
\includegraphics[width=0.75\textwidth]{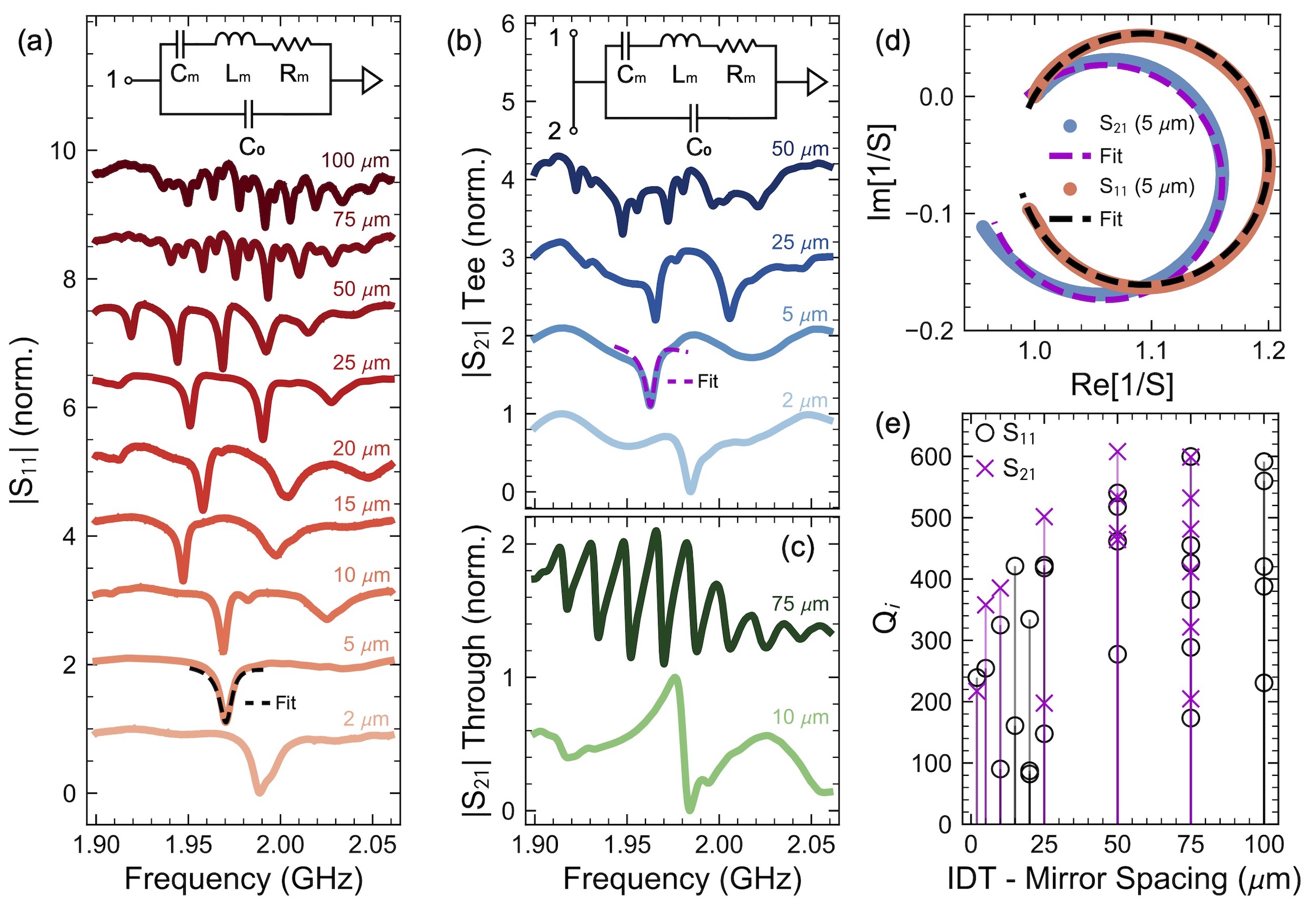}
\caption{
Suspended resonator cavities measured at room temperature.
(a) $|S_{11}|$ reflection measurement using the configuration shown inset in the panel equivalent circuit, for different IDT-mirror spacings. (b) $|S_{21}|$ transmission measurements with the configuration shown inset in the panel equivalent circuit, for different IDT-mirror spacings. (c) $|S_{21}|$ transmission measurements with the configuration shown inset in Fig.\ref{fig:1}(b), for different IDT-mirror spacings. All spectra are normalized to the absolute value of their main dip and vertically offset for clarity.
(d) Imaginary part of the inverse scattering parameter $1/S$ for the measurements in panels (a) and (b), for two resonator cavities with IDT–mirror spacing of 5~$\mu$m, together with the corresponding fits. The scattering parameter subscripts $11$ and $21$ are omitted. The normalized magnitude of these fits are shown as dashed lines in panels (a) and (b) for the corresponding spectra.
(e) Fit intrinsic quality factors $Q_i$ for all measured cavities with different IDT–mirror spacings, obtained from both reflection and transmission measurements. When multiple $Q_i$ values are shown for a given IDT–mirror spacing, these correspond to distinct resonant dips of a single resonator cavity.
}
\label{fig:2}
\end{figure*}
 
To gain insight into the spatial profile of the resonant acoustic mode, we also performed finite element method simulations of the resonator cavities. Figure~\ref{fig:1}c shows the simulation model, approximating the cavity in Fig.~\ref{fig:1}a. Here, to reduce computational effort, the acoustic mirrors are limited to 50 fingers. The resonant peak is identified by calculating the admittance on the IDT fingers. Figure~\ref{fig:1}d shows the spectra of the real and imaginary components of the admittance. The resonance occurs at approximately 2 GHz, in good agreement with experiment. Figure~\ref{fig:1}e shows the displacement field associated with the resonant mode. In the upper part of the panel, we show a top view of the cavity, illustrating how the acoustic mirrors confine the resonant mode at the center, overlaying the displacement amplitude with outlines of the IDT and mirrors; the linear color scale is inset bottom right. 
Note that the aperture is insufficient to fully contain the mode, leading to partial radial leakage from the cavity and thus to diffraction losses.
In the bottom part, we show a cross-sectional cut in the $xy$ plane through the central axis of the resonant cavity, revealing the expected antisymmetric $A_0$ mode Lamb-wave displacement.

We investigated the room-temperature acoustic response of several resonator cavities with different IDT–mirror spacings. Figure~\ref{fig:2}a shows the $|S_{11}|$ spectra, where one set of IDT electrodes is grounded and the other is connected for a VNA single-port reflection measurement. Each spectrum is first shifted to set its baseline to zero and then normalized to the amplitude of its deepest dip. This normalization facilitates visual comparison of the vertically offset spectra.
Figures \ref{fig:2}b and \ref{fig:2}c display $|S_{21}|$ normalized transmission measurements for devices connected in ``tee'' and ``through'' configurations. In the tee configuration, one set of IDT electrodes is connected to the transmission line between VNA ports 1 and 2 while keeping the other grounded; the through configuration was described previously. We indicate the corresponding IDT–mirror spacing next to each measurement, which are displaced vertically for clarity. As expected, increasing this spacing results in a smaller acoustic Fabry-Perot resonance spacing with a resultingly larger number of resonant dips.

\begin{figure*}[htb]
\centering
\includegraphics[width=0.9\textwidth]{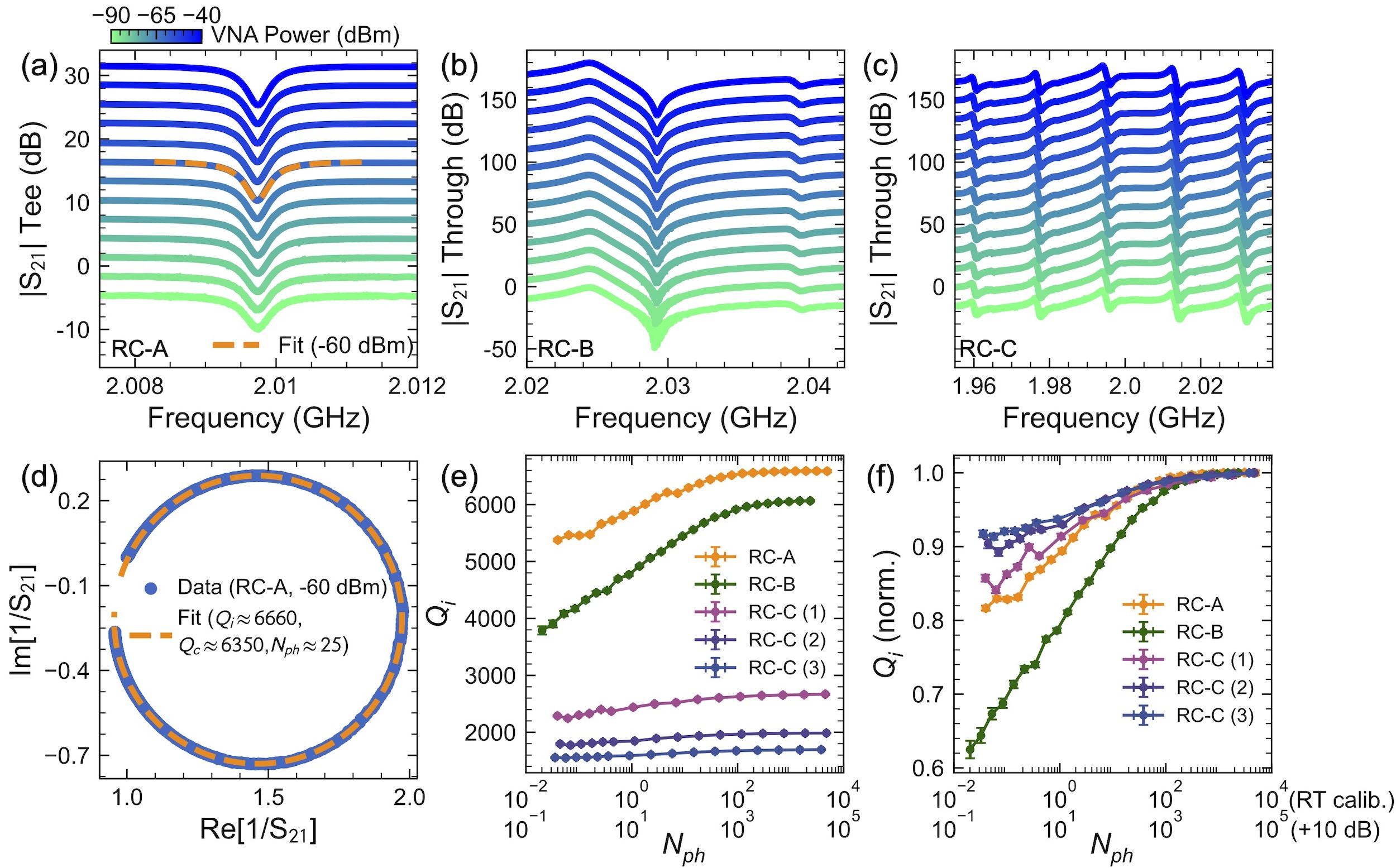}
\caption{
Suspended resonator cavities measured in the millikelvin regime.
(a–c) Transmission spectra for resonator cavity RC-A (tee configuration, panel a), RC-B (through configuration, panel b) and RC-C (through configuration, panel c) measured at $\sim$10~mK. Spectra are vertically offset for clarity. Measurement power (color scale) corresponds to the signal output from VNA, which is heavily attenuated in the cryostat.
(d) Inverse $1/S_{21}$ signal in the complex plane for RC-A (tee measurement; see Table \ref{table:Devices}) with a VNA power of -60~dBm, together with the corresponding fit. The magnitude of the fit is shown in panel (a) superposed on the corresponding spectrum.
(e) Fitted intrinsic quality factors of modes in each of the three cavities as a function of phonon occupation. For RC-A and RC-B, only a single resonance is present, whereas for RC-C, the $Q_i$ for three resonances is displayed.
(f) Same data as in (e), with the quality factors normalized to their respective values at the highest input power. For both (e) and (f) panels, the power scale on the bottom axis is from the VNA power combined with an estimate of the line attenuation (see discussion), with a roughly 10~dB uncertainty in the actual attenuation.
}
\label{fig:3}
\end{figure*}

We fit the complex response of the resonance dips in each spectrum to extract the effective intrinsic $Q_i$ and coupling $Q_c$. For this scope, the measured signal is first normalized by a weakly frequency-dependent global complex background $A(f)e^{ i \theta (f)}$ introduced by the measurement setup \cite{megrant2012planar}. The resulting decoupled signal is then fitted using \cite{megrant2012planar, magnusson2015surface}:
\begin{equation}\label{eq:S11-S21}
\begin{split}
& S_{11} = 1- \frac{2 Q_i/Q_c}{(Q_i+Q_c)/Q_c + i2Q_i \delta f /f_0}e^{i \phi} \\
& S_{21}^{-1} = 1 + \frac{Q_i/Q_c}{1+i2Q_i\delta f /f_0}e^{i \phi},
\end{split}
\end{equation}
where $f_0$ is the resonant frequency, $\delta f = f-f_0$ the detuning, $\phi$ a phase angle introduced by stray reactances. Figure~\ref{fig:2}d shows examples of fits in the complex plane for example $S_{11}$ and $S_{21}$ (tee) measurements, corresponding to the modes of resonator cavities with an IDT–mirror spacing of 5~$\mu$m. The magnitudes of these fit responses are plotted on the corresponding experimental spectra in panels (a) and (b). 

The fitted $Q_i$ from all the different resonator cavities are shown in Fig.~\ref{fig:2}e as a function of the IDT–mirror spacing. Resonator cavities with larger IDT–mirror spacing have a greater metal-free area, which appears correlated with higher $Q_i$, although saturating for larger spacings, possibly due to beam-steering limitations.

\subsection{Millikelvin regime}

\begin{table}[htb]
\centering
  \caption{Suspended resonator cavities measured at $\sim~10$~mK, with fit intrinsic and coupling quality factors $Q_i$ and $Q_c$ at a VNA power of –60~dBm.}
  \label{table:Devices}
  \begin{tabular}{ccccc}
  \hline
    Device & IDT-mirror dist. & Config. & $Q_i$ & $Q_c$ \\
        \hline
        RC-A & 5~$\mu$m & Tee & 6600 & 6350 \\
        RC-B & 10~$\mu$m & Through & 5900 & 670 \\
        RC-C (1) & 75~$\mu$m & Through & 2610 & 660 \\
        RC-C (2) & 75~$\mu$m & Through & 1960 & 940 \\
        RC-C (2) & 75~$\mu$m & Through & 1690 & 740 \\
    
    \hline
  \end{tabular}
\end{table}

\begin{figure*}[thb]
\centering
\includegraphics[width=0.9\textwidth]{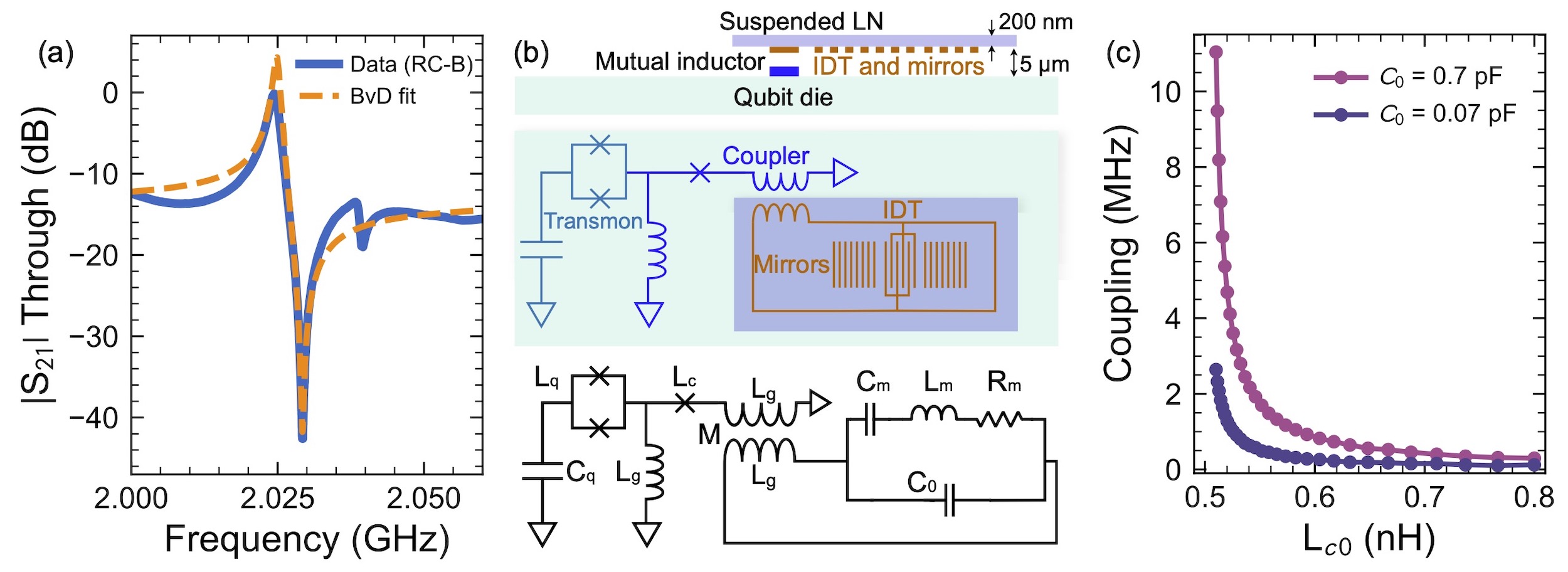}
\caption{
Coupling between a transmon qubit and a suspended resonator cavity.
(a) Transmission spectrum $|S_{21}|$ in the millikelvin regime for RC-B and corresponding BvD fit.
(b) Schematic illustration and equivalent circuit diagram using the BvD model for a transmon qubit inductively coupled to a suspended resonator cavity.
(c) Calculated coupling as a function of the coupler junction inductance $L_{c0}$ for a suspended resonator cavity with an aperture of 10~$\mu$m ($C_0=0.07$~pF) and of 100~$\mu$m ($C_0=0.7$~pF).
}
\label{fig:4}
\end{figure*}

The devices shown in Table \ref{table:Devices} were mounted on the mixing chamber plate of a dilution refrigerator and measured at the base temperature of approximately 10~mK. We completed VNA measurements of three resonator cavities: RC-A was measured in a tee configuration, with an IDT–mirror spacing of 5~$\mu$m; RC-B and RC-C were measured in a through configuration with IDT–mirror spacings of 10 $\mu$m and 75~$\mu$m, respectively. Figures \ref{fig:3}a–c show the $|S_{21}|$ spectra measured at different VNA powers for the three resonator cavities. RC-A and RC-B, with short IDT–mirror spacings, exhibit a single dominant resonance, whereas RC-C displays multiple resonances. For all resonator cavities, we fit the spectra in the complex plane using the same normalization procedure as in Fig.~\ref{fig:3}d and Eq.~\ref{eq:S11-S21} to extract the intrinsic and coupling $Q$-factors, and estimate the average phonon occupation number, $N_{ph}$, in the cavity at a given input power. We estimate the occupation number using the formula \cite{Satzinger2018quantumthesis}
\begin{equation}\label{eq:Nph}
N_{ph} = \frac{2Q_c}{2 \pi f_0} \big(\frac{Q_i}{Q_i+Q_c} \big)^2 \frac{P_{in}}{\hbar 2\pi f_0},
\end{equation}
where $P_{in}$ is the power delivered to the resonator cavity, estimated by accounting for the attenuation in the input line of the power provided by the VNA. The attenuation of each input line was measured at room temperature prior to device installation.
As the attenuation was measured at room temperature without the devices connected, this estimate may be off by as much as 10~dB.

Figure~\ref{fig:3}d shows a parametric fit for $S_{21}(f)$ measured for RC-A at a VNA power of –60~dBm, yielding $Q_i\approx 6660$, $Q_c\approx 6350$, $N_{ph}\approx 25$ (based on the room-temperature attenuation calibration). This fit is also displayed as a dashed orange line in panel (a) superposed with the corresponding measurement. Table \ref{table:Devices} reports the fit results at –60~dBm VNA power for all devices, with three resonances fit for RC-C.

By repeating the fit at different powers, we can extract the dependence of $Q_i$ on the excitation power, reported in Fig.~\ref{fig:3}e and, in normalized form, in Fig.~\ref{fig:3}f, as a function of the estimated phonon occupation $N_{ph}$. The horizontal axis reflects two estimates for the phonon occupation, the upper scale using the room temperature calibration and the lower scaled adding an estimated additional 10~dB loss at low temperatures, which is perhaps a more realistic estimate.

We observe two main features. First,  the internal quality factor $Q_i$ is higher for resonator cavities with shorter IDT–mirror spacings, suggesting better mode confinement compared with longer cavities when material losses are reduced at millikelvin temperatures, although admittedly the sample set is very small and cavity length not the only variable. Second, $Q_i$ increases with the phonon occupancy and saturates around $N_{ph} \approx 100$ (based on the room-temperature cable attenuation measurement). This can be attributed to losses induced by coupling between the resonator and two-level systems in the material \cite{megrant2012planar,Manenti2016,Luschmann2023,GruenkeFreudenstein2025}. At higher powers, these two-level systems become saturated, reducing their impact on $Q_i$.

\subsection{Designs for coupling to superconducting transmon qubits}

From the cryogenic characterization of the RC-B device in the through configuration, we use the BvD model to extract the lumped-element parameters of the Lamb-wave mode at around 2 GHz. Figure~\ref{fig:4}a shows the resonant peak together with the corresponding BvD fit. We use this to create a tunable inductive coupling architecture that connects a suspended lithium niobate resonator to a transmon qubit using a flip-chip geometry \cite{Satzinger2018, Satzinger2019}. As shown in the top and middle panels of Fig.\ref{fig:4}b, a suspended lithium niobate resonator on its own die can be flip-chip coupled to a low-loss, non-piezoelectric sapphire or silicon die where high-quality superconducting transmon qubits can be fabricated with standard techniques. The coupling between the qubit and the resonator is mediated through a mutual inductance with no galvanic connection. In Table \ref{table:scq} we display a design circuit together with the BvD model parameters. This coupling scheme allows tunable coupling by changing the coupler Josephson inductance $L_c=L_{c0}/\cos{\delta}$, with $\delta$ the phase difference across the coupler junction. Maximum coupling between the qubit and acoustic resonator is achieved when the coupler inductance $L_c=-L_{c0}$ \cite{Chen2014}. In Fig.\ref{fig:4}(c), we calculate the maximum achievable coupling when the qubit is brought into resonance with the acoustic resonator, as a function of the coupler inductance $L_{c0}$, using designs with two different IDT apertures (geometric capacitances $C_0$).

\begin{table}[htb]
\centering
  \caption{Design parameters for coupling acoustic resonator to a transmon qubit, both for an acoustic resonator as measured (RC-B) and for an alternative design with a larger aperture.}
  \label{table:scq}
  \begin{tabular}{ccc}
  \hline
    Parameter & Measured & Alternative\\
        \hline
        $C_0$ & 0.07~pF & 0.7~pF\\
        $C_m$ & 0.303~fF & 3.03~fF\\
        $L_m$ & 20.4~$\mu H$ & 2.04~$\mu H$\\
        $R_m$ & 29~$\Omega$ & NA\\
        qubit capacitance $C_q$ & 90~pF & 90~pF\\
        flip-chip mutual $M$ & 0.15~nH & 0.15~nH\\
        ground inductor $L_g$  & 0.2~nH & 0.2~nH\\
        coupler wiring inductance $L_{w}$ & 0.1~nH & 0.1~nH\\
    \hline
  \end{tabular}
\end{table}


\section{Conclusion}
We report on the fabrication and characterization of mechanically-suspended acoustic resonator cavities defined on a thin lithium niobate plate. We show that the fundamental resonant mode observed in these devices is compatible with a gigahertz-frequency Lamb mode in the lithium niobate layer, in reasonable agreement with finite element simulations. We investigate resonators with different IDT-mirrors spacings, with measurements at both room and millikelvin temperatures. The devices exhibit different leakage channels in the two temperature regimes, as discussed in the text. Resonator cavities with small IDT-mirror spacings exhibit internal quality factors as high as approximately 6000, with some dependence of the internal quality factor on drive strength. Using a BvD model, we extract the relevant equivalent circuit parameters to enable evaluation of coupling such cavities to a transmon qubit, with achievable coupling strengths larger than 1 MHz. Couplings as large as 10 MHz should be achievable \cite{Satzinger2018, ArrangoizArriola2019}.

\begin{acknowledgments}
This work is supported by the Japan Science and Technology Agency ASPIRE project (JPMJAP2316), by the Japan Science and Technology Agency Moonshot R\&D grant (JPMJMS2062), and by the JSPS KAKENHI WAKATE (25K17905). We made us of the Institute of Industrial Science (The University of Tokyo) Clean Room Facilities. We also acknowledge support from the Air Force Office of Scientific Research (AFOSR grant FA9550-20-1-0270 and MURI grant FA9550-23-1-0338), DARPA DSO (grant HR0011-24-9-0364), and in part by UChicago's MRSEC (NSF award DMR-2011854), by the NSF QLCI for HQAN (NSF award 2016136), by the Simons Foundation (award 5099), a 2024 Department of Defense Vannevar Bush Faculty Fellowship (ONR N000142512032), and the Army Research Office and Laboratory for Physical Sciences (ARO grant W911NF2310077). Results are in part based on work supported by the U.S. Department of Energy Office of Science National Quantum Information Science Research Center Q-NEXT. We made use of the Pritzker Nanofabrication Facility, which receives support from SHyNE, a node of the National Science Foundation's National Nanotechnology Coordinated Infrastructure (NSF Grant No. NNCI ECCS-2025633). The authors declare no competing financial interests. Correspondence and requests for materials should be addressed to A. N. Cleland (anc@uchicago.edu), M. Nomura (nomura@iis.u-tokyo.ac.jp), or M. Diego (diego@iis.u-tokyo.ac.jp).
\end{acknowledgments}

\appendix
\section{Fabrication details}
\label{app:A}
All devices were fabricated from a commercial Y-cut lithium niobate–on–insulator wafer, consisting of a 200-nm-thick LiNbO$_3$ layer on a silicon dioxide–on–silicon substrate.
The resonator cavities were patterned by electron-beam lithography, followed by the deposition of a 40-nm-thick aluminum layer and a lift-off process. The IDT fingers are designed to launch acoustic waves along the $x$-axis of the lithium niobate.
Contact pads were defined by photolithography, followed by deposition of a 160-nm-thick aluminum layer and subsequent lift-off. Circular windows were patterned by photolithography and etched through the lithium niobate plate using Ar ion milling. Finally, the silicon dioxide sacrificial layer was removed using a hydrofluoric acid vapor etching process to release the structures.

\section{Simulation details}
\label{app:B}
The finite element method simulations were carried out in the COMSOL Multiphysics environment, using the Solid Mechanics and Electrostatics modules coupled via the piezoelectric multiphysics interface. Simulations were performed in the frequency domain. 

Within each pair of fingers forming the IDT, a voltage was applied to one finger while the other was kept grounded. The voltage was modulated at the driving frequency, allowing for the retrieval of the piezoelectric response in the resonator cavity. Free mechanical boundary conditions were applied to the top and bottom surfaces of the lithium niobate layer, as well as to all aluminum surfaces, except at the interfaces between lithium niobate and aluminum, where displacement continuity is assumed. The perimeter of the suspended structure was laterally surrounded by a perfectly matched layer to prevent artificial reflections of outgoing acoustic waves.

\section{Room temperature measurements}
\label{app:C}
Room-temperature reflection measurements were performed using a VNA connected to the device through a calibrated microwave probe station. Transmission measurements were conducted by connecting the VNA ports to a printed circuit board on which the sample was wire-bonded. In all cases, both the magnitude and phase of the signal were recorded, which is essential for the complex-plane fitting discussed in the main text.

\section{Millikelvin measurements}
\label{app:D}

\begin{figure}[htb]
\centering
\includegraphics[width=0.4\textwidth]{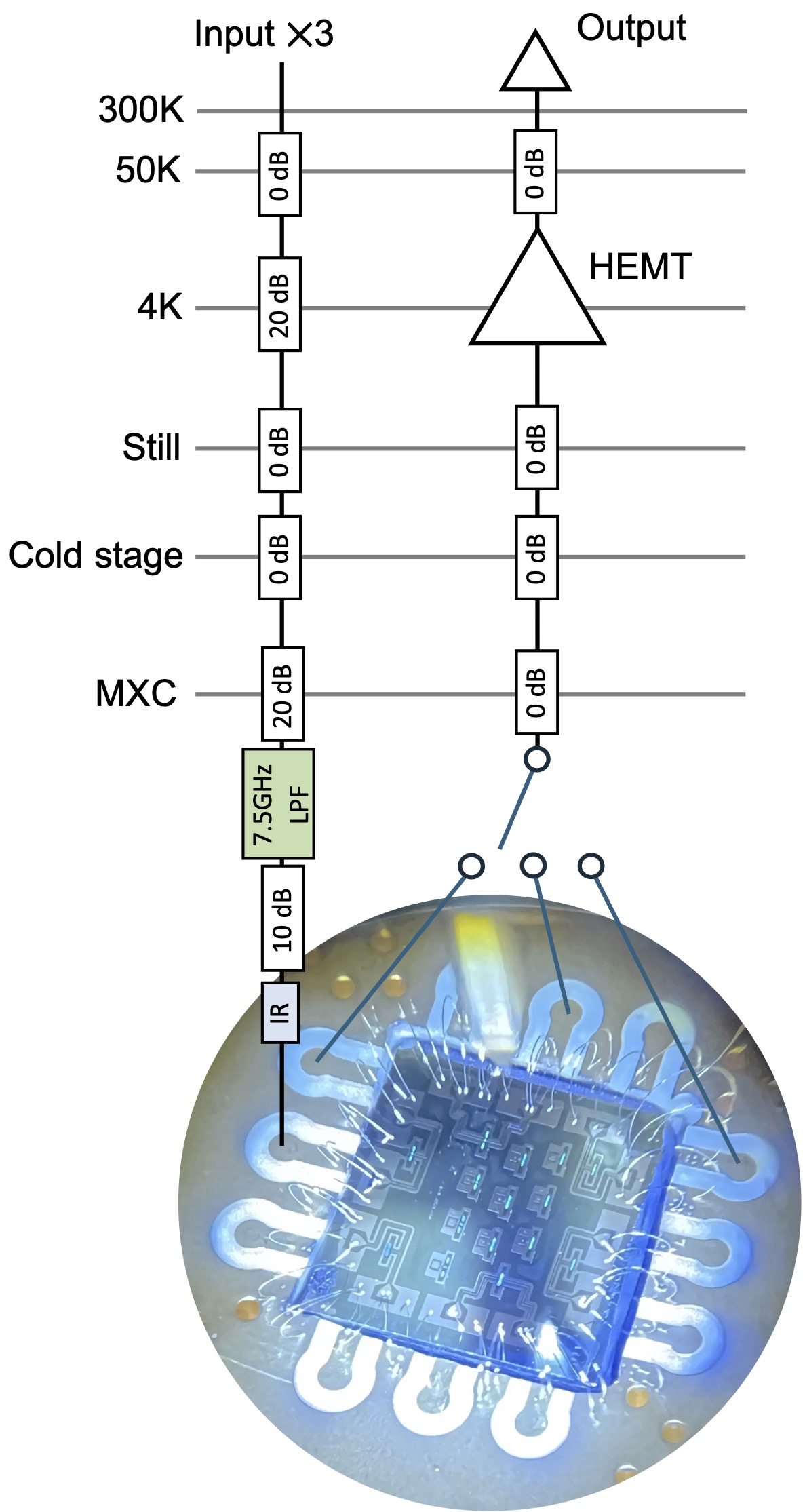}
\caption{
Schematic of the circuitry in the dilution refrigerator, together with a photograph of the sample wire-bonded to the printed circuit board.
}
\label{fig:5Apx}
\end{figure}

Millikelvin-temperature transmission measurements were performed in a configuration similar to that used at room temperature, with the addition of attenuation introduced by the circuitry inside the dilution refrigerator. Figure~\ref{fig:5Apx} shows a schematic of the electrical configuration in the dilution refrigerator. Three identical input lines were connected to the different resonator cavities, using a common output line. A microwave switch allows selection of the resonator cavity.

The attenuation along each input line in the dilution refrigerator was characterized at room temperature. The attenuation at 2 GHz was measured to be –66~dB for the line connected to RC-A, –64~dB for RC-B, and –60.5~dB for RC-C. The output signal from the device was routed through an output line with an in-line room-temperature amplifier.

\bibliography{apssamp}

\end{document}